# The dynamics of amplified spontaneous emission in CdSe/ZnSe quantum dots

D. O. Kundys,[a)] P. Murzyn, J.-P. R. Wells, A. I. Tartakovskii, and M. S. Skolnick
*Department of Physics and Astronomy, University of Sheffield, Sheffield S3 7RH, United Kingdom*

Le Si Dang
*Laboratoire de Spectrométrie Physique (CNRS UMR 5588), Université Joseph Fourier, Grenoble BP 87, F-38402 Saint Martin d'Hères Cedex, France*

E. V. Lutsenko and N. P. Tarasuk
*Stepanov Institute of Physics of NASB, Independence Avenue 68, 220072 Minsk, Belarus*

O. G. Lyublinskaya, A. A. Toropov, and S. V. Ivanov
*Ioffe Physico-Technical Institute of RAS, Politekhnicheskaya 26, St. Petersburg 194021, Russia*

We have used the variable stripe technique and pump-probe spectroscopy to investigate both gain and the dynamics of amplified spontaneous emission from CdSe quantum dot structures. We have found modal gain coefficients of 75 and 32 cm$^{-1}$ for asymmetric and symmetric waveguide structures, respectively. Amplified spontaneous emission decay times of 150 and 300 ps and carrier capture times of 15 and 40 ps were measured for the structures with high and low material gains respectively. The difference in the capture times are related to the fact that for the symmetric waveguide, carriers diffuse into the active region from the uppermost ZnMgSSe cladding layer, yielding a longer rise time for the pump-probe signals for this sample.

## I. INTRODUCTION

CdSe quantum dots have attracted considerable attention for the potential realization of green laser diodes[1–3] (LDs) for applications in laser display technology and short range fibre optic communication links. Previous studies[4–6] have shown that electrically pumped CdSe/ZnSe quantum dot (QD) lasers usually face several degradation problems mainly caused by the presence of point defects, which diffuse from the *p*-type doped upper layers into the active region. Optically pumped LDs do not suffer from such degradation and therefore the idea of obtaining an integrated chip containing a CdSe QD laser structure pumped by a InGaN/GaN blue quantum well laser diode looks very promising.[7]

Most of the time-resolved measurements on self-assembled CdSe QDs, have focused on time resolved photoluminescence where the dynamics of trions, biexcitons, and excitons have been studied.[8] Typical photoluminescence decay times of ~0.5–1 ns are measured for excitons, with a trion lifetime comparable to that of the exciton while the biexciton lifetime was found to be shorter, typically by a factor of 2 compared to that of the exciton. Trion binding energies of 15–22 meV and biexciton binding energies of 19–26 meV were found for dots grown by molecular beam epitaxy (MBE). Further studies have focused on lateral carrier transfer in $Cd_xZn_{1-x}Se/ZnS_ySe_{1-y}$ quantum dot layers.[9] The observed photoluminescence transients had a slow decay component of the order of a few nanoseconds which was attributed to lateral carrier migration within the $Cd_xZn_{1-x}Se$ QD layer eventually leading to the formation of excitons in single QDs.

The dynamics of amplified spontaneous emission (ASE) has been studied for near infrared InAs/GaAs QD lasers.[10,11] However, relatively little is known about the dynamics of ASE for CdSe/ZnSe QDs. Understanding of the emission dynamics should provide important information for the design of QDs suitable for both CW and high-frequency laser applications.[12]

## II. SAMPLES AND EXPERIMENTAL DETAILS

The CdSe / ZnSe QD structures studied in this report were grown on GaAs (001) substrates using a MBE setup equipped with both II-VI and III-V chambers, allowing growth of CdSe/ZnSe heterostructures on top of the deposited GaAs buffer layers. CdSe QDs with a nominal thickness in the range of 2–3 ML (monolayers) were formed by the modified Migration Enhanced Epitaxy (MEE) technique, which uses a very slow deposition rate of 0.3 ML of CdSe per cycle.[13] The layer structures of the two samples studied are shown in Fig. 1. The samples are cleaved pieces of planar wafers. The sample design was optimized in order to minimize substrate losses and maximize the optical confinement factor $\Gamma$. As the result, samples A and B have confinement factors of 0.0185 and 0.0206, respectively. The mean dot size is 4 – 6 nm and the array density is approximately $(1-5) \times 10^{11}$ dots/cm$^2$ for both samples. The samples consist of (i) ZnMgSSe cladding layers, (ii) ZnSSe / ZnSe strain compensated short period superlattice waveguide core regions, and (iii) an array of CdSe QDs placed in the ZnSe quantum well as an active region.

[a)] Corresponding e-mail: Dmytro.kundys@manchester.ac.uk

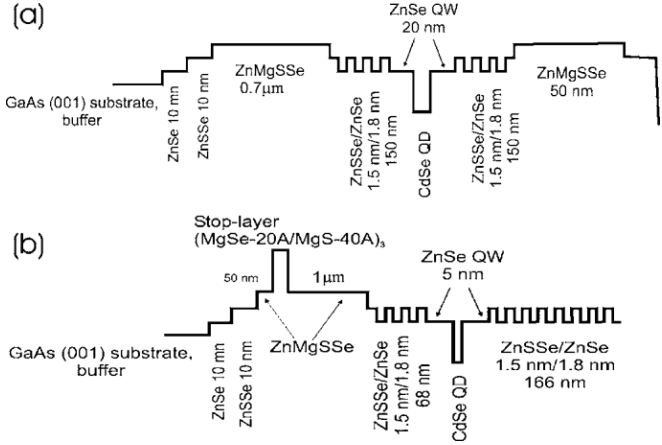

FIG. 1. Conduction band profile of the samples studied. Here and in the rest of the figure captions (a) and (b) represent samples A and B, respectively.

The asymmetric structure of the wave- guide in sample B was designed in order to contrast with symmetric structure A to optimize the confinement factor, and to study the contribution of carriers excited in the upper- most cladding layer which diffuse into the active region from the outermost cladding in sample A. The "stop layer" in sample B is designed to improve carrier localization in the active region (Fig. 1).

Cathodoluminescence (CL) was measured at room temperature, as a first optical characterization of the laser structures. It was found that, whatever the electron acceleration voltage between 1 and 20 kV, the QD layer emission is always dominant over that of the thick layers in the heterostructures, e.g., the ZnMgSSe buffer layer or the ZnSSe/ZnSe waveguide superlattice. This is illustrated in Fig. 2 for an acceleration voltage of 10 kV. At this voltage value, electron-hole pairs are generated at a depth of about 600–700 nm, extending well beyond the active QD layer. Nevertheless the CL of the QDs is dominant by more than two orders of magnitude, demonstrating the high sample quality which allows the diffusion of most generated carriers in the buffer layer and the superlattice and their subsequent capture in the QDs. It is worth noticing that the CL spectra of the two samples are quite similar, suggesting a similar overall sample quality.

Two approaches were used to study the carrier-photon dynamics of optically pumped edge-emitting QD laser structures. Optical gain measurements were made using the variable stripe technique,[14] while a wavelength degenerate pump-probe technique was employed to study the dynamics of the ASE of the QD ground state transition. In order to perform these measurements, a regeneratively amplified, femtosecond Ti:sapphire laser was used to pump a TOPAS optical parametric amplifier (OPA). The OPA has a tuning range from 240 nm to 18 μm and yields output pulses of the order of 100 fs at a repetition rate of 1 kHz. The excitation wavelength was set to 380 nm (3.26 eV) in order to excite above the ZnMgSSe cladding layer. This selection of the excitation wavelength is based on the cathodoluminescence data where the band-edge emissions originating from ZnMgSSe and ZnSeSe were found at 2.87 and 2.72 eV, respectively. Both pump and probe beams were focused with a cylindrical lens giving stripe dimensions close to 2 mm in length and 200 f.Lm in width on the sample surface. The stripe length could be varied by moving an adjustable razor blade mounted on a one-dimensional translation stage in front of the sample. The emitted light from the edge of the sample was collected and coupled into a spectrometer and then dispersed onto a liquid nitrogen cooled charged coupled device (CCD) for spectral analysis. All measurements were carried out at $T=10$ K using an Oxford Instruments "Optistat."

### III. RESULTS AND DISCUSSION

Figure 3 shows the emission spectra for various stripe lengths at a fixed excitation density of 32 f.LJ/cm$^2$/pulse. The spectra are dominated by the inhomogeneously broadened QD ground state emission having a full width at half maximum (FWHM) of 13 meV which is centred around 505 nm for sample A [Fig. 3(a)]. For sample B, the FWHM is about 16 meV with the emission peak centred near 500 nm [Fig. 3(b)]. As the stripe length increases, a strong nonlinear increase in emission intensity is seen due to the occurrence of ASE. The inset shows a plot of the emission intensity versus stripe length. The slope of the straight line fraction of the plot (see insets to Fig. 3) yields the modal gain coefficient $g$ from a fit to

$$I \propto e^{gl} - 1 \qquad (1)$$

As the stripe length increases further, gain saturation occurs.[15] In Fig. 3 it occurs at stripe lengths of 0.8 and 1.3 mm for the symmetric and asymmetric structures (samples A and B), respectively. The gain saturation indicates that the number of photons in the cavity is so large that a considerable number of them interact with QDs already deexcited by other photons, thus producing no further amplification. The nonlinearities in sample B occur at a longer stripe length (around 0.7 mm) than in sample A. The fit to the exponential part of the curves yields a high modal gain of $g=32$ cm$^{-1}$ for sample A and a value which is almost twice as large for sample B of $g=75$ cm$^{-1}$. These modal gains are very high, particularly by comparison with the well studied InAs QD laser systems, where modal gains of ~6 cm$^{-1}$ are typical. The most likely origin of the high modal gains are the high dot densities achievable in the II-VI system, about one order of magnitude higher than those for InAs–GaAs QDs.[16] One can see that the difference in confinement factors between the two samples is not significant enough to cause such a big difference in their modal gain values, suggesting that the enhancement in the modal gain for sample B is influenced by the variation in the sample design. The emission linewidth is narrower and ASE threshold which occurs at shorter stripe length for sample A (symmetric waveguide) is lower as compared to the asymmetric waveguide, sample B. This feature is most likely attributed to the lower internal losses for the symmetric waveguide type sample.

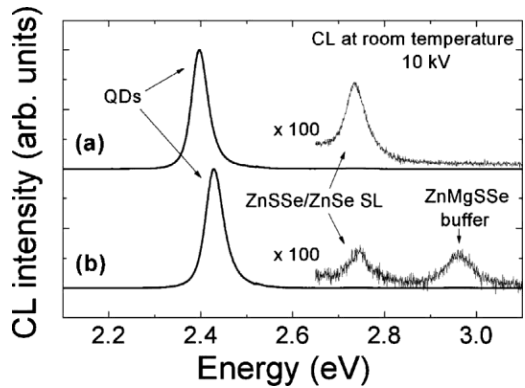

FIG. 2. Room temperature cathodoluminescence measurements of the studied samples.

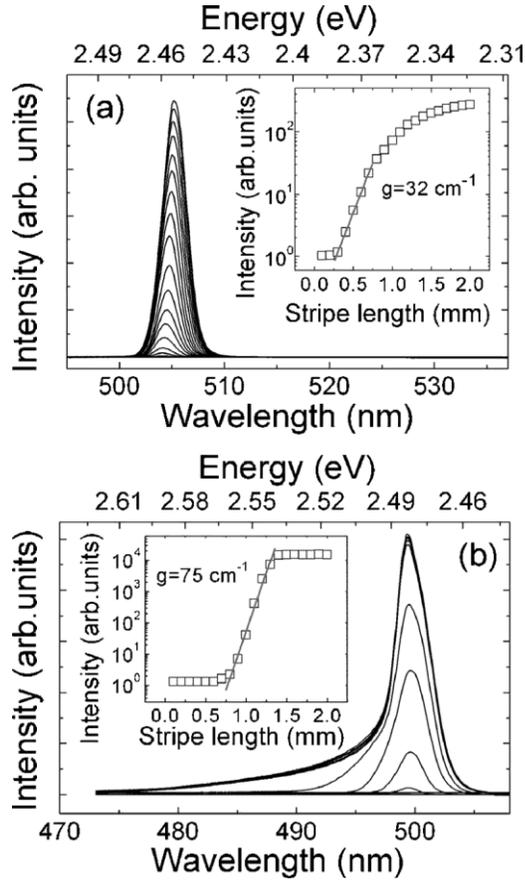

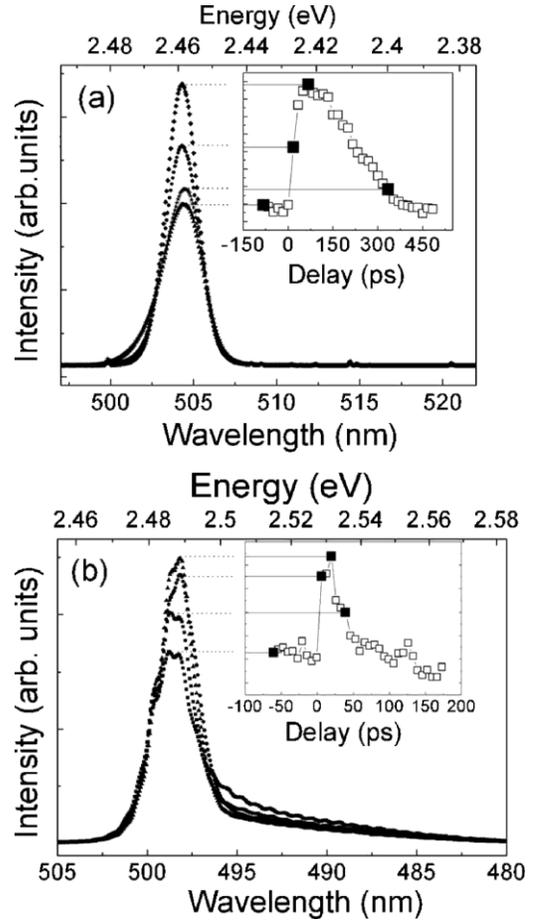

FIG. 3. ASE spectra for different stripe lengths measured at 10 K and 32 f.LJ/cm² excitation energy density for samples A and B, respectively. The insets show the maximum intensity as a function of stripe length; the solid line is a fit to Eq. (1).

FIG. 4. Time-resolved emission spectra from QDs laser structure of samples A and B. The insets demonstrate evolution of the pump-probe emission intensity as a function of time delay.

A study of the ASE dynamics was performed using a wavelength degenerate pump-probe technique with two colinear beams exciting the sample from the top surface. A strong pump pulse excites electron-hole pairs in both the ZnMgSSe and ZnSSe barriers. These electron-hole pairs are subsequently captured by the QDs where recombination takes place. At high pump powers a nonlinear increase of the ASE is observed. A weaker probe pulse creates additional carriers at a controlled time delay after the pump. Depending on the pump-probe delay, the additional carriers excited by the probe lead to an enhancement of the ASE (Ref. 11). Therefore, by changing the delay between the pump and probe pulses, the dynamics of the ASE from the QD structures can be measured, providing information on the QD optical pumping cycle in both the time and spectral domains. The pump and probe beam powers were selected in a range suitable to measure the strongest nonlinearity in the ASE.

For sample A, the pump and probe pulse energies were selected at 80 and 10 nJ corresponding to excitation densities of 20 and 2.5 μJ/cm², respectively. The results of the pump-probe measurements are shown in Fig. 4(a). The emission induced by the probe alone was about 15 times smaller than that of the pump. At the same time additional excitation by the probe together with the pump induces a very large (up to 40%) increase in the emission demonstrating a clear nonlinear dependence of the ASE at times following the pump pulse. The pump-probe transient measured for sample A exhibits a decay time of ~350 ps [Figure 4(a)] and a rise time of ~40 ps associated with carrier capture. A 3 meV emission linewidth narrowing is also observed when the delay between the pulses is changed from negative values to zero. This nonlinear pump-induced linewidth narrowing is due to the onset of ASE. The decay of 350 ps is due to recombination of electron hole pairs and appears to be faster than the exciton recombination rates measured from the time-resolved photoluminescence measurements for the MBE grown self-assembled QDs where decay times of 500–1000 ps were measured (Ref. 8).

The pump-probe results for sample B are shown in Fig.

4(b) for pump and probe excitation densities of 18 and 1.7 μJ/cm$^2$, respectively. The decay time of the ASE for sample B is ~150 ps [inset to Fig. 4(b)]. The decrease of the decay time can be explained in terms of the higher gain coefficient in this sample, following from the equation,[17]

$$g = \frac{R_r}{v_g N_p}(f_2 - f_1), \quad (2)$$

where $v_g$ and $N_p$ are the group velocity and photon density, respectively, $R_r$ represents the radiative transition rate that would exist if all states were available to participate in the transition, and $f_i$ is the Fermi distribution function. Thus for a fixed excitation density, the radiative transition rate in the amplified regime is proportional to the gain coefficient which is consistent with our data. A 3 meV linewidth narrowing was also observed for sample B when comparing the $t=0$ and negative time delay data. The modulation observable in Fig. 4(b) is due to Fabry-Pérot interference fringes. A high resolution pump-probe scan yields a rise time of about 15 ps which is of the same order as the calculated finite photon lifetime in the cavity of the studied length. This is due to the different capture mechanisms from sample A. In that case, the excited carriers in the top ZnMgSe cap layer need to diffuse into the active region from the uppermost ZnMgSSe cladding layer, thus leading to the longer lifetime seen for this sample.

## IV. CONCLUSION

In summary, a study of the low temperature dynamics of ASE from CdSe/ZnSe QD laser structures has been carried out. Gain coefficients of 32 and 75 cm$^{-1}$ have been obtained for symmetric and asymmetric waveguide structures respectively, which were determined at an excitation wavelength of 380 nm and for an excitation density of 32 μJ/cm$^2$. Measured modal gain values are about 5 times higher than those previously reported for III-V based QD laser structures. This finding is attributed to the variations in the sample designs and probably the structural quality and QD density. A pump-probe technique has been employed to study the dynamics of the ASE where the capture times are resolvable and decay times of 150 and 350 ps were measured which can be explained in terms of the higher gain coefficient measured for the asymmetric structure.